\documentclass[aps,prl,showpacs,psfig,epsfig,twocolumn]{revtex4}

\usepackage[dvips]{graphicx}
\input{epsf}
\begin{document}
%\draft
\title{AC field induced quantum rectification effect in tunnel junctions}
\author{M. V. Fistul}
\affiliation{Physikalisches Institut III,
Universit\"at Erlangen-N\"urnberg, D-91058, Erlangen, Germany}
\author{A. E. Miroshnichenko}
\affiliation{$^2$Max-Planck-Institut f\"ur Physik komplexer Systeme, N\"othnitzer
Strasse 38, D-01187 Dresden, Germany }
\author{S. Flach}
\affiliation{$^2$Max-Planck-Institut f\"ur Physik komplexer Systeme, N\"othnitzer
Strasse 38, D-01187 Dresden, Germany }
%\address{Max-Planck-Institut f\"ur Physik komplexer Systeme, N\"othnitzer
%Strasse 38, D-01187 Dresden, Germany }
\date{\today}
%\wideabs{
%\maketitle
\begin{abstract}
We study the appearance of directed current in tunnel junctions 
({\it quantum ratchet effect}), in the presence of an external ac field $f(t)$.
The current is established in a one-dimensional discrete inhomogeneous
"tight-binding model". 
By making use of a symmetry analysis we predict the right choice of $f(t)$ 
and obtain the directed current as a difference between electron transmission 
coefficients in opposite directions, $\Delta T~=~T^{LR}-T^{RL}$.
Numerical simulations confirm the predictions of the symmetry analysis and  moreover, 
show that the directed current can be drastically increased by a proper choice of 
frequency and amplitudes of the ac field $f(t)$.
\end{abstract}

\pacs{05.60.Gg, 73.23.Ad, 73.40.Gk}

\maketitle
%}

A great attention has been devoted to theoretical and experimental studies of 
a transport rectification in various physical, chemical and biological systems 
\cite{Prost,Reinm}.
In particular this peculiar effect appears in the form of a directed current as a system 
is exposed to a time-dependent ac force with zero mean. 
Such a directed motion (current) has been observed in 
molecular motors \cite{Prost,Julicher}, 
in Josephson junction coupled systems \cite{Josmix,MooijQuant}, and in systems of 
cold Rb atoms in laser fields \cite{Schiavoni}, just to name a few.

In most studies the appearance of directed current has been analyzed by making use of 
the model of a particle moving in the presence of both space and time periodic potentials
\cite{Reinm,Hangi,FistPRE,FlachZolEvt}.
The effects of dissipation and interaction with a heat bath 
have been taken into account, and in many cases the decrease of dissipation as a system 
approaches the Hamiltonian limit, leads to a substantial increase of the directed 
current \cite{FlachZolEvt}. 
By making use of a symmetry analysis of {\it the dynamic equations of motion} 
it has been shown that the directed current occurs as a space-periodic 
potential is asymmetric (ratchet potential) and/or all important symmetries of an ac force, 
i.e. time-reversal and 
shift symmetry, are broken \cite{Reinm,FlachZolEvt}. Notice here, that the latter case can be 
easier realized in experiments. 
It has been established that more complex systems described by generic 
Bolzman transport equation \cite{FlachZolEvtepl}, Fokker-Planck equation 
\cite{Hangi,FistPRE,FlachOvch}, etc. also display a directed current.
Moreover, the directed energy transport has been obtained in the case of interacting 
many particle systems described by nonlinear partial differential equations 
\cite{FFMZprl,SolZol}.

Most previous studies analyzing both the dissipative and Hamiltonian limits, 
were heavily based on a {\it classical } nonlinear regime of particle motion. A next 
step is to obtain the directed transport in systems displaying {\it quantum-mechanical} 
behavior. 
The quantum regime of directed transport for a few particular systems has been studied 
in Refs. \cite{MooijQuant,Goych,Vinokur,Hangi2}. It was shown that a dissipationless quantum system 
in a single energy band tight-binding approximation does not support directed current \cite{Goych}, 
and although a quantum Brownian particle can display a directed motion and even numerous 
current reversal in the presence 
of an asymmetric ratchet potential, one still needs to take into account the dissipative 
effects \cite{Vinokur}. Notice that a quantum ratchet effect has been observed 
in Josephson junction arrays \cite{MooijQuant} as the directed motion of a Josephson vortex 
in a ratchet potential with two interacting energy bands.   

On the other hand there is a class of intrinsically quantum phenomena
involving a particle tunneling through a potential barrier where effects 
connected with the dissipation in tunnel region can be diminished. The particular examples of systems 
displaying these phenomena are the ballistic regime in mesoscopic devices, 
various tunnel junctions, tunneling through quantum (molecular) wires, etc.
Recently the directed current has been obtained for electrons that tunnel 
through the barrier containing the asymmetrically distributed energy levels in the 
presence of harmonic ac drive \cite{Hangi2,Hangi3}. However, in this case the dissipation caused 
by the coupling to the leads is crucial in order to obtain a nonzero directed current. 
Thus, it is interesting to obtain general 
conditions for the appearance of directed transport in tunnel (dissipationless in the tunnel region) 
systems that are subject to an externally applied ac drive.  

In this Letter we consider electron tunneling through a one-dimensional potential 
barrier $U(x)$ in the presence of an externally applied ac force $f(t)$. Moreover, we will restrict 
ourselves to the {\it elastic} scattering case, i. e. the energy of incoming electron coincides 
with the energy of transmitted electrons.
By making use of symmetry arguments we obtain the conditions  for the directed 
current in such a system and verify these arguments by direct numerical simulations. 
We show how the directed current can be resonantly enhanced by a proper 
choice of the frequency and amplitudes of ac drive. Notice here, that the obtained current 
can be considered as a particular example of the widely discussed quantum pump effect 
\cite{Qpump1,Qpump2,Buttiker}.

In the absence of dissipation the propagation of electrons through the potential barrier $U(x)$ 
($U(x)_{x \rightarrow \pm \infty} \rightarrow 0$) is described by the one-dimensional 
Schr\"odinger equation.
The externally applied ac force $f(t)$ induces a time-dependent potential that is odd in space, 
i.e.  $-xf(t)$.
The crucial characteristics of electron propagation
is the transmission coefficient $T (E) $ which 
in the elastic case depends on the energy of electron $E$. Moreover, we will distinguish the electron 
transmission coefficients in the opposite directions: $T^{LR}(E)$ as the 
electron moves from left to right 
and $T^{RL}(E)$ as the electron moves from right to left. 
Thus, the directed current in such a system appears if a symmetry is broken and the {\it difference }
$\Delta T(E)~=~T^{LR}(E)-T^{RL}(E)$ is not zero. 
The symmetry properties of the transmission coefficient $T(E)$ can be analyzed by making use of the 
Green function method \cite{Econ} and especially its representation in the form of path integral 
\cite{Feynm}.
Indeed, the transmission coefficient $T(E)$ is written in the form:
\begin{equation} \label{GreenF}
T(E)~\propto~\int_{-\infty}^{\infty} dt e^{\frac{-iEt}{\hbar}} G(x_1,x_2;t)~~,
\end{equation}
where the coordinates $x_1$ and $x_2$ are located on the different sides of the barrier. 
Thus the symmetry of transmission coefficients $T^{LR (RL)}(E)$ and correspondingly the absence of 
directed current is determined by a symmetry of the Green function $G(x_1,x_2;t)$ with respect to 
the permutation of $x_1$ and $x_2$. In order to obtain the condition allowing a nonzero 
directed current we represent the Green function in the form of the path integral \cite{Feynm}:
$$
G(x_1,x_2;t)~=~\int D[u(t)] e^{-\frac{iS(u(t);~x_1,x_2)}{\hbar}}~=~
$$
\begin{equation} \label{PathInt}
 ~=~\int D[u(\tau)] e^{-\frac{i}{\hbar}\int_{-\infty}^\tau d\tau 
 \frac{m_{eff}}{2}(\frac{du}{d\tau})^2-U(u(\tau))+u(\tau)f(\tau)}~~.
\end{equation}
Here, $m$ is the electron effective mass, and all paths $u(\tau)$ start from the coordinate $x_1$ at 
$\tau~=~-\infty$ and pass the point with the coordinate $x_2$ at $\tau~=~t$.
A most important symmetry transformation is {\it time reversal}. We notice that time reversal 
$\tau ~\rightarrow~ -\tau$ changes (permutes) the coordinates $x_1$ and $x_2$, and the Green function 
$G(x_1,x_2;t)$ is invariant with respect to this permutation and correspondingly the directed 
current is absent, provided $f(t)$ is symmetric \cite{comment}:
\begin{equation} \label{SymmCond1}
 f(t)~=~f(-t)
\end{equation}
A simple consequence of such a time-reversal symmetry is that the directed current vanishes for an 
arbitrary potential $U(x)$ in the absence of an ac force.

Next, we consider another symmetry operation: {\it reflection in space}. 
In this case the symmetry operation $x \rightarrow -x$, and correspondingly $u \rightarrow -u$ 
in the integral of Eq. (\ref{PathInt}), changes the coordinates $x_1$ and $x_2$ but 
leaves the Green function $G(x_1,x_2;t)$ invariant, provided the potential $U(x)$ is symmetric 
and the ac force is {\it shift symmetric}:
\begin{equation} \label{SymmCond2}
 U(x)~=~U(-x);~~~f(t)~=~-f(t+\frac{ \pi }{\omega})~~,
\end{equation}
where $\omega$ is the period of ac force.

It is interesting to mention that the conditions (\ref{SymmCond1}) and (\ref{SymmCond2}) of the absence 
of directed transport obtained for 
the quantum-mechanical system, coincide with ones obtained for classical dynamical systems in 
Hamiltonian limit \cite{FlachZolEvt,FlachZolEvtepl,FFMZprl}.
Thus, a simplest choice to obtain the directed current is to use the symmetric potential 
$U(x)$ and the ac force $f(t)$ 
allowing to break the shift symmetry and time-reversal symmetry 
\begin{equation} \label{ftcond}
 f(t)~=~ f_1 \sin \omega t + f_2 \sin(2\omega t +\varphi).
\end{equation}
Note that only for the phase shift $\varphi~=~\pm \pi/2$ the ac force $f(t)$ possesses 
the time-reversal symmetry, and the directed current has to vanish.

To verify the above presented  analysis and also to study the dependence of the directed current 
on the frequency and amplitudes of the ac drive, we consider a particular example of a tunnel junction, 
namely a large and wide barrier between 
two narrow-band electrodes (Fig. 1). 
The ac force of the type (\ref{ftcond}) is applied to the barrier
region. In a simplest case this system can be described by a standard tight-binding model in the presence of 
a time-dependent scattering potential:
\begin{equation} \label{equationNUM}
i\hbar \dot \psi_n~=~c_0^2 (2\psi_{n}-\psi_{n+1}-\psi_{n-1})+U_n\psi_{n}-f_n(t)\psi_{n}~~.
\end{equation}
The potential $U_n$ and $f_n(t)$ have the form:
\begin{equation} \label{Potential}
U_n ~=~ \left\{
\begin{array}{rl}
U_0~~, &  -2 \leq ~n~ \leq 2 \nonumber \\  
0~~, & n~ <~-2;~~ n~> ~2.
\end{array}
\right.
\end{equation}
\begin{equation} \label{PotentialTimeDep}
f_n (t)~=~ \left\{
\begin{array}{rl}
-n f(t)~~, &  -2 \leq ~n~ \leq 2 \nonumber \\  
0~~, & n~ <~-2;~~ n~> ~2.
\end{array}
\right.
\end{equation}
For simplicity  
the hopping between the sites determined by $c_0$, is 
assumed to be the same in the electrodes and in 
the barrier region. Note here, that this model can be directly used also for the analysis of electron 
propagation through the periodic chain of impurities with symmetrical (see, Fig. 1) 
or asymmetrical distribution of energy levels \cite{Hangi2,Hangi3}.

\begin{figure}
%\vspace{20pt}
\includegraphics[scale=0.3]{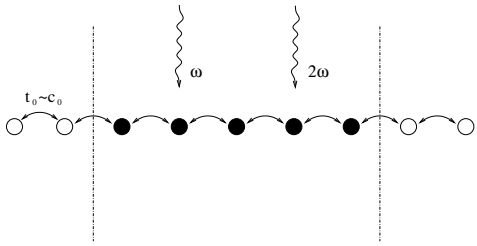}
%\centerline{\psfig{figure=fig1-schem1.eps,width=82mm,height=60mm}}
%\vspace{2pt}
\caption{The schematic view of a system (tunnel junction) 
allowing to obtain a quantum-mechanical directed current. 
Dashed lines show the boundary of a tunnel barrier region. The tunnel junction is modelled by 
a chain of different atoms (black circles). 
The electron hopping between atoms is the same in electrodes and in the barrier region, 
but the energy levels
of atoms in the barrier region are shifted and form a tunnel barrier.
The ac force is applied to the 
barrier region only. 
}
%\label{fig1}
\end{figure}

We use  $c_0~=~0.1$ to ensure that the case of elastic scattering
(a one channel scattering) was realized. The height of the barrier was fixed $U_0~=~1$ and thus, 
in the absence of ac drive the transmission coefficient was negligibly small ($~\simeq~10^{-12}$).
In order to {\it resonantly enhance} the transmission coefficient and the directed current 
the frequency $\omega~=~0.5$ was used. For this particular value of $\omega$ both harmonics in 
(\ref{ftcond}) 
give a contribution to the resonant enhancement. 
Such an enhancement of the transmission coefficient is due to a well known photon assisted tunneling. 
As the ac drive is applied, 
the electron can absorb one or two photons, then propagate through the barrier region with the enhanced 
energy $\tilde E~=~E+2\hbar \omega ~\geq~U_0$ and emit the 
phonons at the end 
of this region. The transmission coefficient $T(E)$ for such a process can be large and 
moreover, can be different for electrons propagating in different directions. 

To calculate the transmission coefficient we turn to a numerical investigation.
We developed a scheme that generalizes the one given in \cite{FCret}, and allows to compute the 
transmission coefficient for any time-periodic scattering potential. The details of this method 
can be found in \cite{FFMtr}, and this scheme has been used in order to calculate the phonon 
scattering by time-periodic and localized in space breathers \cite{FFMtr}.

The dependence of the computed transmission coefficient $T^{LR}(q)$ on the wave vector $q$ of 
electron propagating from left to 
right is shown in Fig. 2. Notice here that in our model 
the wave vector $q$ is determined by a standard 
relationship: $E~=~4c_0^2 \sin^2(q/2)$. 
As  expected the transmission coefficient drastically increases and 
moreover, displays the effect of resonant tunneling for particular values of the energy $E$ of a
propagating electron. For these values of energy the transmission coefficient is close to one. Notice that the 
resonant transmission obtained for the electron wavevector $q~=~\pi/2$
 has a simple physical origin. Indeed, for this particular energy the resonant 
condition $qL~=~2\pi$, where $L$ is the length of the barrier region, is satisfied and the transmission
has a maximum.
As the amplitudes of ac force $f_{1,2}$ or the phase shift $\varphi $ vary
additional resonances appear in the dependence $T^{LR}(q)$ (see Fig. 2). These resonances in 
$T^{LR}(q)$ are due to the resonant propagation of an electron with enhanced energy 
$\tilde E~=E+2\hbar \omega$ through the ac force dependent energy levels of a barrier region.

% Figure 2
%
%----------------------------------------------------------------------
%\begin{figure}[htb]
\begin{figure}
\vspace{20pt}
\includegraphics[scale=0.3]{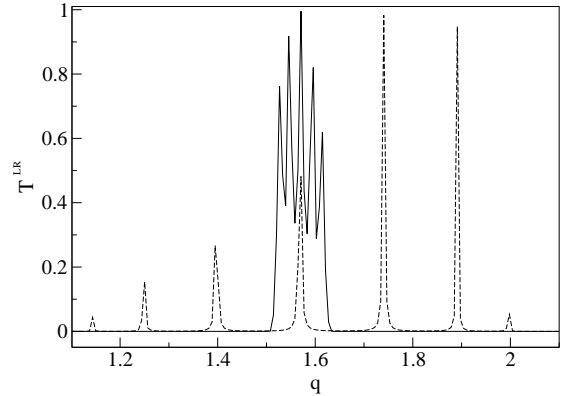}
%\centerline{\psfig{figure=fig1.eps,width=72mm,height=50mm}}
\vspace{2pt}
\caption{The wave vector $q$ dependence of the transmission coefficient $T^{LR}(q)$:
one harmonic case $f_1=0.9$ and $f_2=0$ (solid line), and 
two harmonic case where the amplitudes 
$f_1~=~0.9$ and $f_2~=~1$ were used (dashed line). The value $\varphi=\pi/2$ was used.}
%\label{fig1}
\end{figure}

Moreover, we observed a strong dependence with a sign change of the difference between transmission 
coefficients $\Delta T(E)~=~T^{LR}(E)-T^{RL}(E)$ on the phase shift $\varphi$ (see Fig. 3). As 
it was expected from generic symmetry analysis, $\Delta T(E)$ is practically zero as the phase shift 
$\varphi~=~\pm \pi/2$. 
The value of $\Delta T(E)$ strongly changes with the amplitudes of ac drive, and the maximum 
value of $\Delta T(E)~\simeq~0.043$ (around $7.5 \%$ from the value of $T(E)$) 
has been found for particular values of $f_{1,2}$.

% Figure 3
%
%----------------------------------------------------------------------
%\begin{figure}[htb]
\begin{figure}
\vspace{20pt}
\includegraphics[scale=0.3]{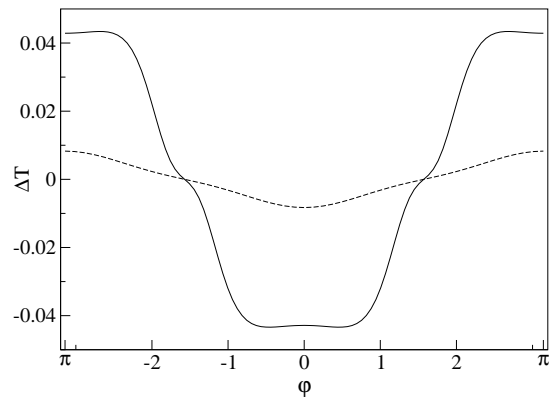}
%\centerline{\psfig{figure=fig2.eps,width=72mm,height=50mm}}
\vspace{2pt}
\caption{The dependence of $\Delta T$ on the phase shift $\varphi$. 
The curves are shown for a particular 
value of wave vector $q~=~\pi/2$ as the resonant transmission occurs. 
Solid line: 
$f_1~=~2.36$ and $f_2~=~-0.03$; dashed line:  $f_1~=~2.36$ and $f_2~=~-0.003$. 
Note here that the latter case is scaled by a factor of 10, so the real values of $\Delta T $ 
is 10 times less than they appear in the plot.}
%\label{fig1}
\end{figure}

Notice, here that an increase of the spatial extension of the ac drive "spot" can lead to a 
substantial decrease of both transmission coefficient and directed current. 
It is due to a specific of our model where a small value of coupling $c_0$ has been chosen. 

The physical origin of the considered rectification effect is the 
quantum-mechanical interference between various paths of an electron
with different absorption (emission) of photons. For this particular case of ac induced
resonant propagation in tunnel junctions, 
this interference is especially transparent. As an electron propagates from left to right it absorbs one or two photons on 
the left side of a junction, and after the propagation, emits the photons on the right side of 
the junction. The amplitude of this process is $A(-f_1^2+if_2e^{i\varphi})(-f_1^2+if_2e^{-i\varphi})$,
where $A$ is the amplitude of electron transmission with an enhanced energy.
Similarly, for the electron propagating from right to left we obtain 
the amplitude $A(-f_1^2-if_2e^{i\varphi})(-f_1^2-if_2e^{-i\varphi})$, and correspondingly, 
the difference in the transmission coefficients occurs, which is proportional to 
$\cos (\varphi )$. 

In order to observe this resonant rectification effect the ac drive of the frequency that is a half 
of the barrier height has to be used. Moreover, the second harmonic also has to be present. Thus, 
it can be obtained with the standard infrared laser sources if the barrier height $U_0$ is around 
$0.1 eV$. 
Our results can be used not only for tunnel junctions but also for the wave propagation through 
various transmission lines, e. g. Josephson junctions arrays, 
in the presence of both a localized static scattering potential $U_0$ and 
an externally applied mixed ac drive of frequencies $\omega~=~U_0/(2\hbar)$ and $2\omega$.

This work was supported by the Deutsche Forschungsgemeinschaft
and the European Union through LOCNET HPRN-CT-1999-00163.
M. V. F. thank the Alexander von Humboldt Stiftung for partial supporting this work.
\bibliographystyle{unsrt}
%\bibliography{bib3}

\begin{thebibliography}{30}
\bibitem{comment} Notice here that our symmetry analysis is based on an assumption of an elastic 
scattering case. In the case of an inelastic scattering even time-reversal ac forces can lead to the 
directed current \cite{Buttiker}. The  importance of time-reversal symmetry for a quantum pump effect 
has been first stressed in Ref. \cite{Buttiker}.
\end{thebibliography}

\end{document}